\documentclass[12pt]{article}
\usepackage{graphicx}
\usepackage{amsmath}
\usepackage{amsthm}
\usepackage{amsfonts}
\usepackage{amssymb}
\usepackage{color}

\title{The Role of Nonlinear Relapse on Contagion Amongst Drinking Communities}

\author{Ariel Cintr\'{o}n-Arias$^{1}$,
Fabio S\'{a}nchez$^2$, Xiaohong Wang$^3$,\\
Carlos Castillo-Chavez$^{3,4,5,6}$, Dennis M. Gorman$^7$, Paul J. Gruenewald$^8$\\\\
\begin{small} $^1$ Center for Research in Scientific Computation \end{small} \\
\begin{small} Box 8205, North Carolina State University, Raleigh, NC 27695-8205 \end{small}\\\\
\begin{small} $^2$Department of Biological Statistics and Computational Biology \end{small}\\
\begin{small}Cornell University, Ithaca, NY 14853-7801\end{small}\\\\
\begin{small}$^3$Mathematical, Computational, and Modeling Sciences Center\end{small}\\
\begin{small}P.O. Box 871904, Arizona State University, Tempe, AZ 85287 \end{small}\\\\
\begin{small}$^4$School of Human Evolution and Social Change\end{small}\\
\begin{small}Arizona State University, Tempe, AZ 85287 \end{small}\\\\
\begin{small}$^5$School of Mathematics \& Statistics \end{small}\\
\begin{small}Arizona State University, Tempe, AZ 85287 \end{small}\\\\
\begin{small}$^6$ Santa Fe Institute\end{small}\\
\begin{small}1399 Hyde Park Road, Santa Fe, NM, 87501 \end{small}\\\\
\begin{small}$^7$ Department of Epidemiology \& Biostatistics, School of Rural Public Health\end{small}\\
\begin{small}Texas A\&M Health Science Center, P.O. Box 1266, College Station, TX 77843-1266\end{small}\\\\
\begin{small}$^8$Prevention Research Center\end{small}\\
\begin{small}1995 University Avenue, Suite 450, Berkeley, CA 94704\end{small}}

\begin{document}
\date{December 4, 2008}
\maketitle{}
\newpage

\begin{abstract}
Relapse, the recurrence of a disorder following a symptomatic remission, is a frequent outcome in substance abuse disorders. Some of our prior results suggested that relapse, in the context of abusive drinking, is likely an ``unbeatable'' force as long as recovered individuals continue to interact in the environments that lead to and/or reinforce the persistence of abusive drinking behaviors.  Our earlier results were obtained via a deterministic model that ignored differences between individuals, that is, in a rather simple ``social'' setting. In this paper, we address the role of relapse on drinking dynamics but use models that incorporate the role of ``chance'', or a high degree of ``social'' heterogeneity, or both. Our focus is primarily on situations where relapse rates are high. We first use a Markov chain model to simulate the effect of relapse on drinking dynamics.  These simulations reinforce the conclusions obtained before, with the usual caveats that arise when the outcomes of deterministic and stochastic models are compared. However, the simulation results generated from stochastic realizations of an ``equivalent" drinking process in populations ``living" in small world networks, parameterized via a disorder parameter $p$, show that there is no social structure within this family capable of reducing the impact of high relapse rates on drinking prevalence, even if we drastically limit the interactions between individuals ($p\approx 0$). Social structure does not matter when it comes to reducing abusive drinking if treatment and education efforts are ineffective. These results support earlier mathematical work on the dynamics of eating disorders and on the spread of the use of illicit drugs. We conclude that the systematic removal of individuals from high risk environments, or the development of programs that limit access or reduce the residence times in such environments (or both approaches combined) may reduce the levels of alcohol abuse.

\end{abstract}




\section{Introduction}\label{intro}

The mechanisms responsible for observed drinking patterns within and between populations
are complex (Daido 2004; Weitzman et al. 2003; Mubayi et al. 2008; and references therein). The development of compartmental and mathematical frameworks geared towards the identification of key ``transition'' mechanisms that increase the percentage of abusive drinkers must factor in the impact of individuals' socioeconomic characteristics, their propensity to drink (heavy drinking tends to run in families), changes in local environments (going to college), treatment failure, ineffectiveness of educational efforts, cultural norms and community values (Mubayi et al. 2008; and references therein).

The term drinking (population) dynamics refers to the study and identification of \lq\lq average" mechanisms, at the individual
level, responsible for observed drinking patterns within the organizational and temporal scales of interest. We model drinking dynamics at the population level as the result of individuals' social contacts in pre-specified environments (\lq\lq drinking contagion").  This modeling approach has proved useful in the identification of the mechanisms behind social patterns that are thought to be, in part, an outcome of intense interactions between individuals in shared social environments.  This modeling approach has been applied to the study of the spread of scientific ideas and innovations (Bettencourt et al. 2006 ); in studies that focus on the mechanisms behind the observed increases in prevalence of eating disorders (Gonz\'{a}lez et al. 2003); in studies that address the impact of relapse on the distribution of drinkers (S\'{a}nchez et al. 2007; S\'{a}nchez 2006); in studies that envision violence as an epidemic (Patten and Arboleda-Florez 2004); as explanation for the observed growth or decline of crime in cities (Gladwell 1996); and in studies that highlight the explosive increases in the use of illicit drugs, such as ecstasy (Song et al. 2006; Mackintosh and Stewart 1979). Researchers are interested in studying the impact of individual drinking habits and preferences' variability at multiple levels of social organization: from small \lq\lq isolated" to highly connected communities; and over short or long time horizons. Models have been used to explore the capacity of  drinking environments to support communities of drinkers as well as the impact of individuals' movements between drinking venues on the overall distribution of drinking types (Mubayi et al. 2008).

The National Institute on Alcohol Abuse and Alcoholism
estimates that 18 million Americans suffer from alcohol abuse or
dependence. Alcohol-related problems cost the United States (U.S.) nearly
$\$$185 billion annually (National Institute of Alcohol Abuse and Alcoholism 2008a) while alcohol abuse was responsible for nearly 80,000 fatalities per year during 2001-05, and it is now the third leading cause of death in the U.S. (Centers for Disease Control and Prevention 2008a).  Prevention
and control efforts that include treatment and education programs that target specific populations including children (Leadership to Keep Children Alcohol Free 2008) or adolescents (College Drinking 2008) are in need of improvement. Among the many problems confronting these programs are the very high rates of relapse after treatment that are observed. Up to 70\% of treated alcohol abusers relapse after treatment (reviewed in S\'{a}nchez et al. 2007). Mathematical studies can be particularly effective as guides to the evaluation, testing and implementation of single or multiple intervention strategies over short or long time scales. This is particularly true in the study of chronic relapsing diseases such as alcohol addiction.

\textbf{Social dynamics, disease transmission, and social structure}\\
Several aspects linked to disease transmission depend strongly on a population's social dynamics. Disease dynamics can often be driven by factors that include heterogeneity in behavior, frequency of use of mass transportation, travel patterns, and cultural norms and practices.  Examples
where the use of mathematical models have generated useful insights include studies on the role of behavior on the transmission dynamics of sexually transmitted diseases like gonorrhea or HIV (Castillo-Chavez et al. 2003;
Hethcote 2000; Hethcote and Yorke 1984; Anderson and May 1991; and references therein) and studies on the intensity and frequency of
travel on the spread of communicable diseases such as SARS (Chowell et al. 2003; Song et al. 2003) and influenza (Hyman et al. 2003; Chowell et al. 2006a). The most significant study of the role of heterogenous mixing on the transmission dynamics
of gonorrhea was carried out by Hethcote and Yorke (Hethcote and Yorke 1984). These researchers
through their introduction of the concept of core group (outliers in the
distribution of sexually-active individuals) showed that most secondary cases of gonorrhea infections could be traced to the core
(most connected nodes in a network of sexually-active individuals). Furthermore, they showed that focusing surveillance and treatment on core subpopulations resulted in significant reductions in gonorrhea prevalence.  The public health policy at that time wrongly focused on the ``random'' testing of women, a policy derived from data that showed that a large percentage of gonorrhea infected women are indeed
asymptomatic (Hethcote and Yorke 1984; and references therein).

The systematic study of the role of heterogenous social landscapes on disease dynamics
began in direct response to efforts to stop the HIV epidemics.  Efforts to compute explicit mixing matrices (who had interactions with whom)
and to study the impact of sexual preference in the context of HIV transmission intensified (Blythe et al. 1990; Blythe et al. 1995;
Blythe et al. 1991; Busenberg et al. 1989; Busenberg et al. 1991; Hsu 1993; Hsu et al. 1994; Hsu et al. 1996;
Castillo-Chavez et al. 1996; Hethcote 2000;
Anderson and May 1991; Castillo-Chavez 1989; and references therein).

Most recently, efforts to explore disease dynamics in the context of heterogenous (fixed) social network structures
have proved quite fruitful. The study of epidemics on network has increased our understanding of the role of ``social'' heterogeneity on disease dynamics (Newman 2003; and references therein) but the impact of the efforts of the mathematical ``network" community goes beyond the study of epidemics on networks, as is evident from the wealth of applications found in the literature (see Watts and Strogatz 1998; Barabasi and Albert 1999; Newman 2003; Newman et al. 2006; and references therein). There is a body of research that contributes to the characterization and validation of some classes of network structures with data (Meyers et al. 2005); structures whose statistical properties are most often captured via power law distributions (Newman et al. 2006).  The class of best known or more popular models of this type
include small-world (Watts and Strogatz 1998) and scale-free (Barabasi and Albert 1999) networks.

Social network analysis is the result (to a great degree) of major contributions by social scientists (Wasserman and Faust 1994; and references therein). Recent contributions by mathematical scientists (Newman 2003 and references therein; Newman et al. 2006; Watts and Strogatz 1998) have increased interactions between social and mathematical scientists. Applications that make use of specialized network structures include studies of the structure of scientific co-authorship networks (Newman 2003), the organizational structure of committees in the U.S. House of representatives (Porter et al. 2005), the structure of internet networks (Pastor-Satorras and Vespignani 2001), the properties of contact tracing networks for SARS (Meyers et al. 2005), and the nature of sexual partnership networks (Liljeros et al. 2001).  Efforts to study stochastic epidemic and social processes  on networks have also been carried out in the context of homeland security  (Chowell and Castillo-Chavez, 2003 and references therein) and drinking (Braun et al. 2006). Our goal here is ``theoretical'', that is, we focus on the study of drinking on some networks characterized by scaling laws (Newman 2003; and references therein). Specifically, the primary objective is to explore the role of network structure on the distribution of drinkers in communities (small world type) where relapse rates are high.

This manuscript is organized as follows. Section \ref{sdrdet} revisits the results
in (S\'{a}nchez et al. 2007; S\'{a}nchez 2006) on the role of relapse on the distribution of drinking types. Section \ref{stowellm}  introduces the stochastic analog
of the deterministic model to highlight the role of variability in the distribution of
drinking types of Section \ref{sdrdet}. Section \ref{swdrink} simulates \textit{one version} of the stochastic drinking dynamics in a small-world network. Finally, Section \ref{dissc} discusses the role of relapse in these settings.

\section{A Deterministic Contagion Model in Well-mixed Drinking Communities}\label{sdrdet}

In the drinking model formulation proposed in (S\'{a}nchez et al. 2007), the population is divided in three classes: $S(t)$, moderate and occasional drinkers (Centers for Disease Control and Prevention 2008c), $D(t)$, problem or heavy drinkers (Centers for Disease Control and Prevention 2008d; National Institute of Alcohol Abuse and Alcoholism 2008b), and temporarily recovered, $R(t)$.  Table \ref{ode_mdl_vars} presents the definitions used in (S\'{a}nchez et al. 2007) where it is assumed that the population is composed of ``average" individuals that interact at random with
each other.  The proportion of contacts of $S$-individuals with $D$-individuals
per unit of time is therefore proportional to $D/N$ where $N=S+D+R$, denotes the total size of the community.  The progression rate from $S$ to $D$ and the relapse rate from $R$ to $D$ depend on frequency-dependent (random) interactions.

In (S\'{a}nchez et al. 2007) the model is given by the following set of nonlinear differential equations:
\begin{eqnarray}
\label{sys1e1}
\frac{dS}{dt} &=& \mu N - \beta S(t) \frac{D(t)}{N} -\mu S(t),\\
\label{sys1e2}
\frac{dD}{dt} &=& \beta S(t) \frac{D(t)}{N} + \rho R(t) \frac{D(t)}{N} - (\mu+\phi)D(t),\\
\label{sys1e3}
\frac{dR}{dt} &=& \phi D(t) - \rho R(t)\frac{ D(t)}{N} - \mu R(t),\\
\label{sys1e4}
N&=&S(t)+D(t)+R(t),
\end{eqnarray}
where $\beta$ denotes the per-capita effective contact rate (transmission rate), that is,
$\beta SD/N$ denotes the rate of transitions from $S$ to $D$, the result of the frequency-dependent interactions between individuals in the classes $S$
and $D$; $\mu$ denotes the per-capita departure rate from the system;
$\rho$ denotes the per-capita effective relapse rate, that is, $\rho RD/N$ denotes
the rate of transitions from $R$ to $D$, the result of the frequency-dependent
interactions between $R$ and $D$; $\phi$ denotes the per-capita recovery (treatment
or education) rate; and $\mu N$ denotes the total recruitment rate into this homogeneous
social mixing community. It is assumed that all ``recruits'' are $S$-individuals. Hence, we set the $S$-recruitment rate equal to $\mu N$ as it guarantees  constant population size.  The validity of the analysis is therefore tied to a time horizon where changes
in total population size are minimal.

The reproductive number under a treatment/education regime $\phi$ is given by
\begin{equation}\label{rphi}
	\mathcal{R}_{\phi}\equiv\mathcal{R}(\phi)=\frac{\beta}{\mu+\phi}.
\end{equation}
$\mathcal{R}_{\phi}$ is a dimensionless quantity (ratio or number) that can be
interpreted as the number of $D$-individuals \lq\lq generated" in a population
of primarily $S$-individuals sharing a common environment. That is, if we start with $S\approx N$ individuals and introduce a ``typical"
$D$-individual then we expect $\mathcal{R}_{\phi}$ secondary cases generated from the $S$ population per $D$-individual, but only at the start of the ``outbreak''. Hence, $\mathcal{R}_{\phi}>1$  results in an exponentially growing $D$-community if $N$ is large enough.  We also expect that when $\mathcal{R}_{\phi}<1$, the introduction of  $D$-individuals in a population where $S\approx N$ ($N$ large) will not result in the growth and (eventual) establishment of a problem-drinking community ($D$-individuals). The above observations are on target when the rate of relapse is linear, that is, $\rho R$ rather than
$\rho R D/N$.  However, when the relapse rate
is nonlinear, namely, $\rho R D/N$, the outcome is not as ``expected". The outcome depends on the ratios
\begin{eqnarray}
	\mathcal{R}_{\rho} &=&\frac{\rho}{\beta} \left[1-\mathcal{R}(\phi)\right]\\
	\mathcal{R}_{c}&=& \frac{\rho}{\beta}\left[\frac{1}{1+\frac{1}{\mathcal{R}_0}}-2\sqrt{\frac{1}{\mathcal{R}_0}-\frac{\mu}{\rho}}\right],
\end{eqnarray}
where $\mathcal{R}(\phi)$ is defined in Equation (\ref{rphi}); $\mathcal{R}_0\equiv\mathcal{R}(0)=\beta/\mu$.

$\mathcal{R}_{\rho}$ can be interpreted as the number of problem drinkers ($D$-individuals) generated from the $R$-class
as a result of the frequency-dependent interactions between the $R$- and $D$-classes ($R$-individuals remain in the same environment).  We observe that $\mathcal{R}_{\rho}>0$
if and only if $\mathcal{R}(\phi)<1$.  On the other hand $\mathcal{R}_{c}>0$ but only as long as
\[
	\frac{\beta}{\mu+\beta}>2\sqrt{\frac{1}{\mathcal{R}_0}-\frac{\mu}{\rho}}>0.
\]
We have not been able to interpret the meaning of $\mathcal{R}_c$ in social terms. However, the value of $\mathcal{R}_c$, under some
conditions, provides a sharp $D$-extinction threshold, that is, a threshold that if crossed, would lead to the eventual elimination of the $D$-class, independent of initial conditions ($D(0)$).

\begin{figure}[h]
\begin{center}
\includegraphics[width=3in,height=2in]{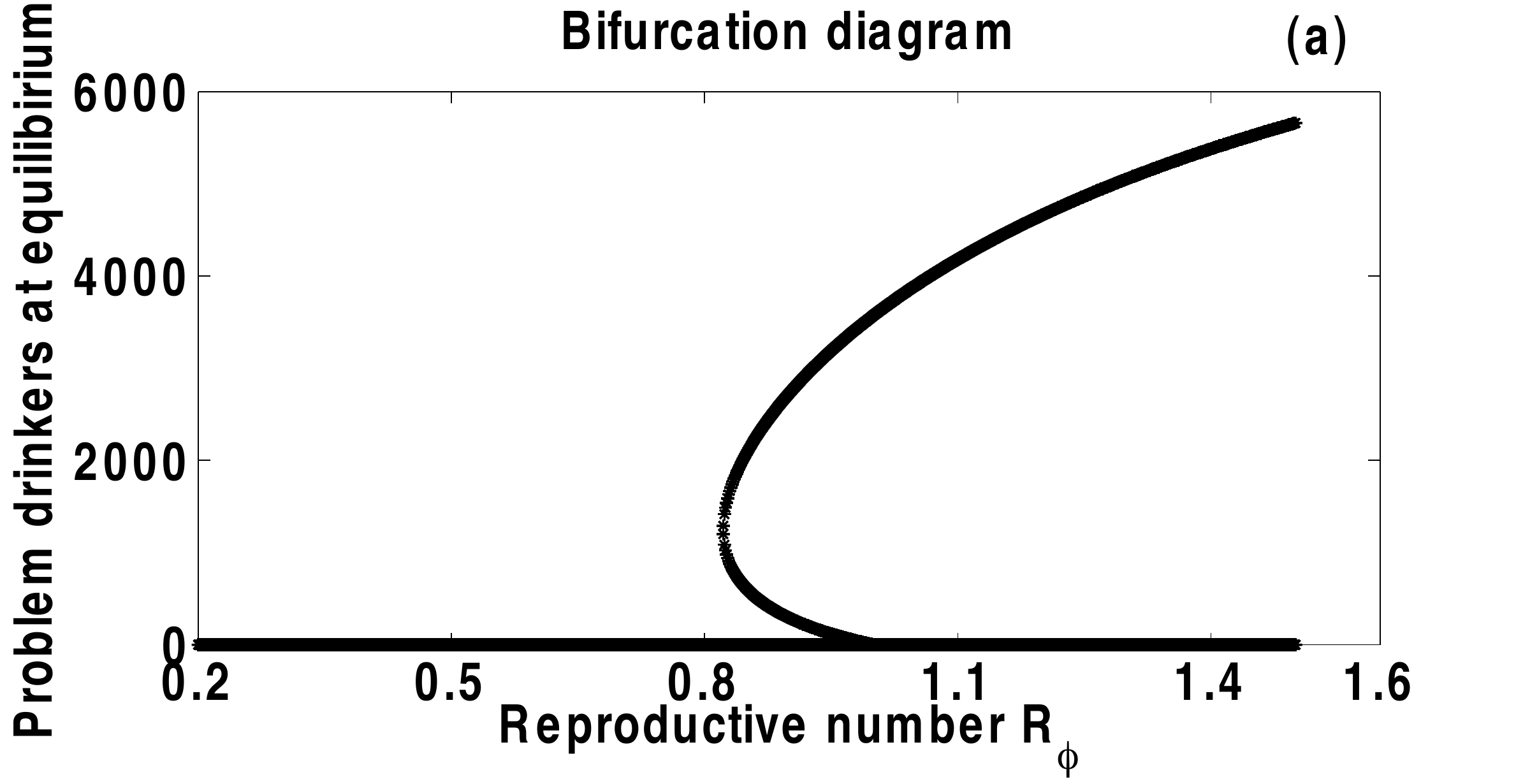}
\includegraphics[width=3in,height=2in]{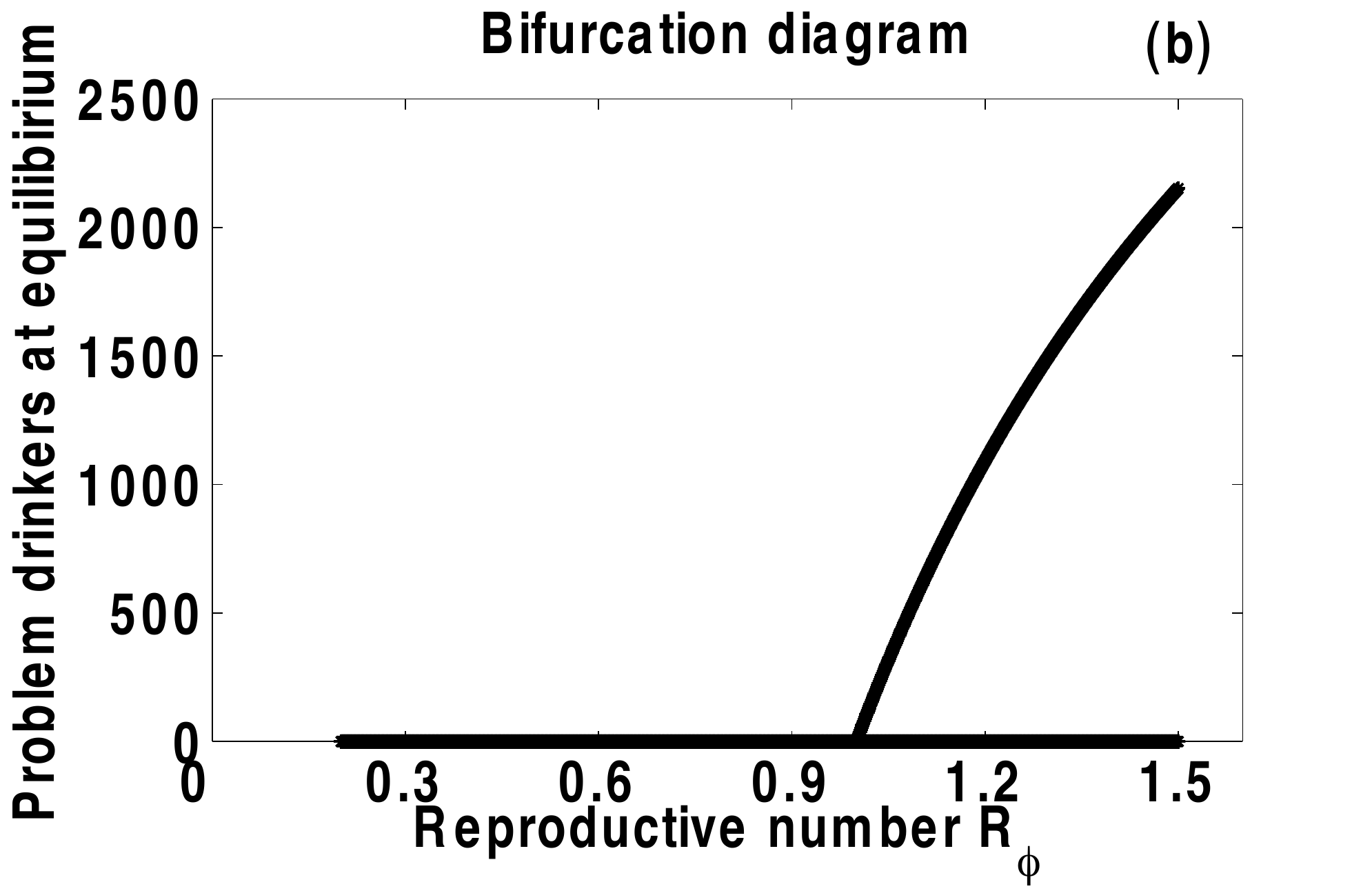}
\includegraphics[width=3in,height=2in]{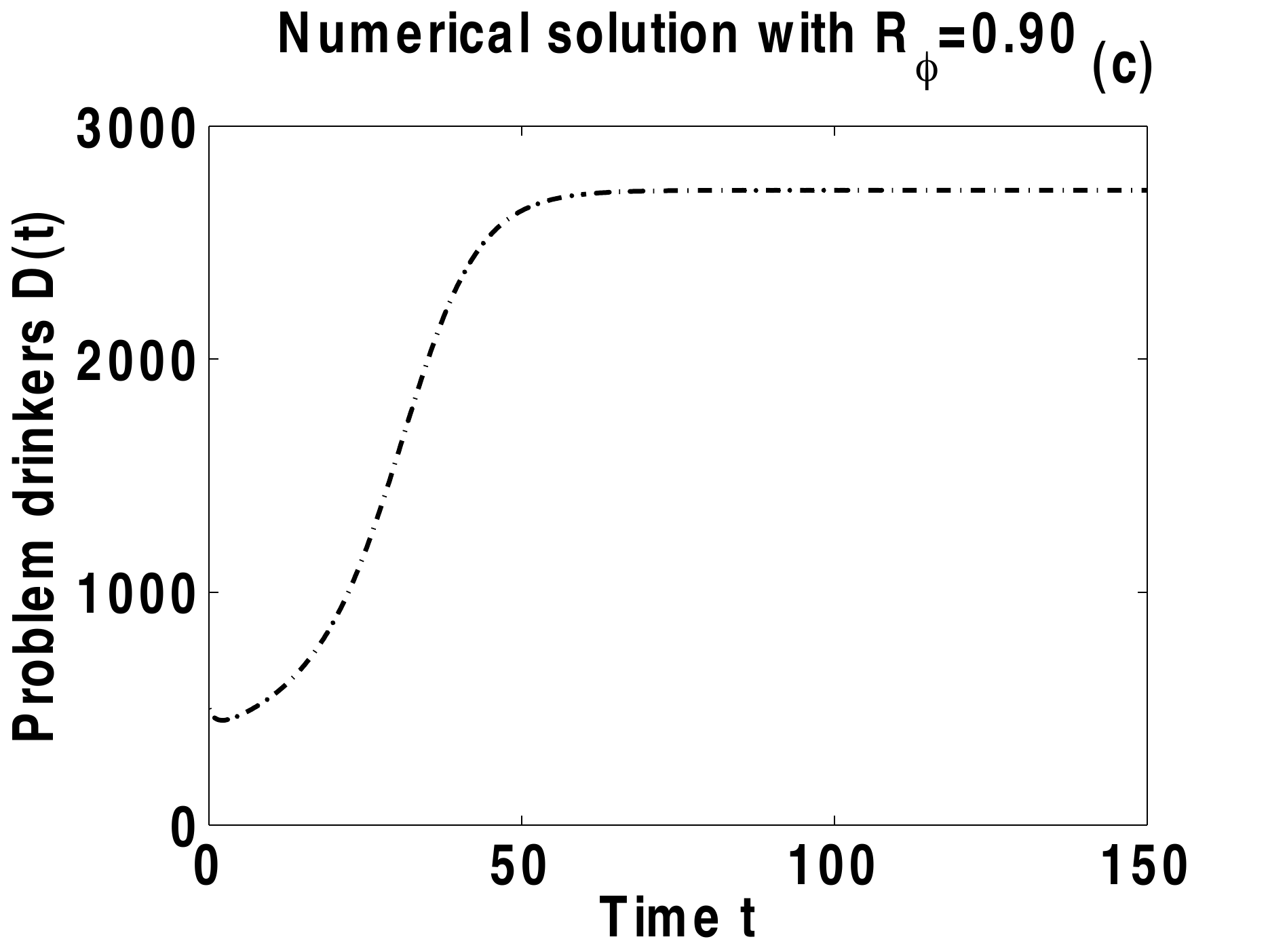}
\includegraphics[width=3in,height=2in]{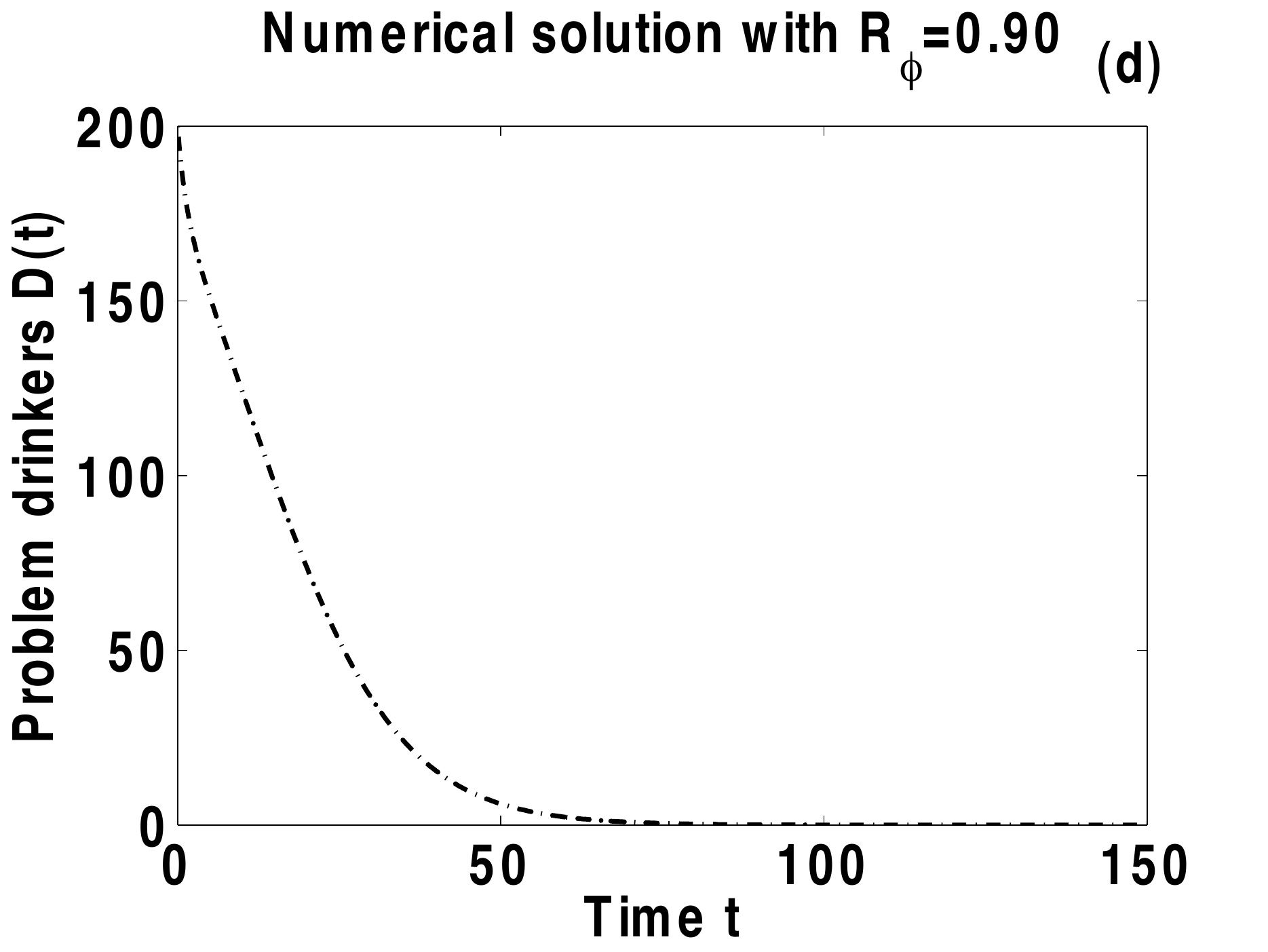}
\caption {Numerical simulations of drinking model in a homogeneous drinking community. Panel (a) shows a bifurcation diagram
that involves the number of problem drinkers at equilibrium versus the reproductive number $\cal{R}_{\phi}$, when $\phi<\rho$. 
Panel (b) displays a bifurcation diagram illustrating the special case when the recovery rate equals the relapse rate
($\phi=\rho=0.50$). Here, ${\cal R}_\phi<1$ provides a sufficient
condition that guarantees the eventual extinction of the population of problem drinkers.  Panels
(c) and (d) display $D(t)$ versus $t$ under different
initial conditions. In Panel (c) the initial conditions are $S(0)=0.98N$, $D(0)=0.02N$ and $R(0)=0$;
in Panel (d) they are $S(0)=0.95N$, $D(0)=0.05N$ and $R(0)=0$. The parameter values used are:
$N=10000$, $\mu=0.50$, $\phi=0.50$ and $\rho=7.00$, \ $0.20\leq\beta\leq1.50$ (Panel (a));
$N=10000$, $\mu=0.50$, $\phi=\rho=0.50$,\ $0.20\leq\beta\leq1.50$ (Panel (b));
$N=10000$, $\mu=0.50$, $\phi=0.50$ and $\rho=7.00$, $\beta=0.90$ (Panels (c) and (d)).}
\label{alcofig}
\end{center}
\end{figure}

The distribution of drinking types, in the nonlinear relapse rate case, depends not only on the thresholds $\mathcal{R}_{\phi}$, $\mathcal{R}_{\rho}$,
and $\mathcal{R}_{0}$ but also on the size of the initial population of problem drinkers, $D(0)$.  In (S\'{a}nchez et al. 2007)
the following results were obtained:
\begin{enumerate}
\item If ${\cal R}(\phi)>1$ then the $D$-class becomes established.
\item Whenever ${\cal R}_c<{\cal R}(\phi)<1$ and ${\cal R}_\rho<1$ or whenever ${\cal R}(\phi)<{\cal R}_c<1$
the $D$-class becomes (eventually) extinct.
\item \label{case3} Whenever ${\cal R}_c<{\cal R}(\phi)<1$ and ${\cal R}_\rho>1$
whether or not the $D$-class becomes established is a function of the initial size of the class of $D$-individuals, $D(0)$
(see Figure \ref{alcofig}(c), (d)).
\end{enumerate}

Numerical simulations ---Figure \ref{alcofig}(a), (c), (d)--- illustrate the role of initial conditions
on drinking dynamics.  Nonlinear relapse leads to a system that supports two socially acceptable coexisting stable
equilibria ($D\equiv0$ and $D>0$). Where the system ends depends on initial conditions. Figures \ref{alcofig}(a), (b) show bifurcation diagrams for the number of problem drinkers at equilibrium as a function of the reproductive number ${\cal R}(\phi)$ (with ${\cal R}_\rho>1$).

A per-capita relapse rate greater than the per-capita recovery rate, $\rho>\phi$, leads to explosive growth in the $D$-class as long as $D(0)$ (the initial population of problem drinkers) is ``large enough'' (see Figure \ref{alcofig}(a)). The qualitative behavior displayed in Figure \ref{alcofig}(a) is commonly called  a ``backward'' bifurcation (S\'{a}nchez et al. 2007). We further observe that once the population of problem drinkers becomes established (${\cal R}_c < {\cal R}(\phi) <1$) their extinction can only be carried out if $\phi$ increases to the point where ${\cal R}(\phi)<{\cal R}_c$ or if $\rho$ decreases to the point where
${\cal R}_{\rho}<1$.  Figures \ref{alcofig}(c), (d), display $D(t)$ versus $t$ to illustrate, with a time series, the effects of initial conditions, $D(0)$. We observe bistability. The size of the initial number of problem drinkers determines whether or not a $D$-community becomes established even under unfavorable conditions (${\cal R}(\phi)<1$).  When the per-capita relapse rate equals the recovery rate, $\rho=\phi$, we observe (Figure \ref{alcofig}(b)) that the $D$-class grows (gradually) with $\mathcal{R}(\phi)$; multiple endemic (non-negative) stable $D$-equilibria will not co-exist in this case.  When $\rho=\phi$, ${\cal R}(\phi)<1$ guarantees the eventual extinction of the problem drinking class.

\section{A Stochastic Contagion Model}\label{stowellm}

The stochastic model of this section is built from the deterministic model given by System (\ref{sys1e1})--(\ref{sys1e4}) and is used to quantify the role of variability on drinking dynamics.  Here, we concentrate on an stochastic analog to the ``mean field'' model given by Equations (\ref{sys1e1})--(\ref{sys1e4}), the deterministic model that supports
two positive equilibria ($\mathcal{R}_c<\mathcal{R}_{\phi}<1$ and $\mathcal{R}_{\rho}>1$).

The derivation of the stochastic model (continuous-time Markov chain) is standard (details are provided in an Appendix)---see for instance (Allen 2003; Allen and van den Driessche, 2006; Renshaw 1991). We carry out simulations that highlight the differences between stochastic and deterministic outcomes. Simulation outcomes (distributions) are later used to contrast the results of stochastic simulations of the same drinking process in small-world networks.
	
\begin{figure}
    \begin{center}
        \includegraphics[width=4in,height=3in]{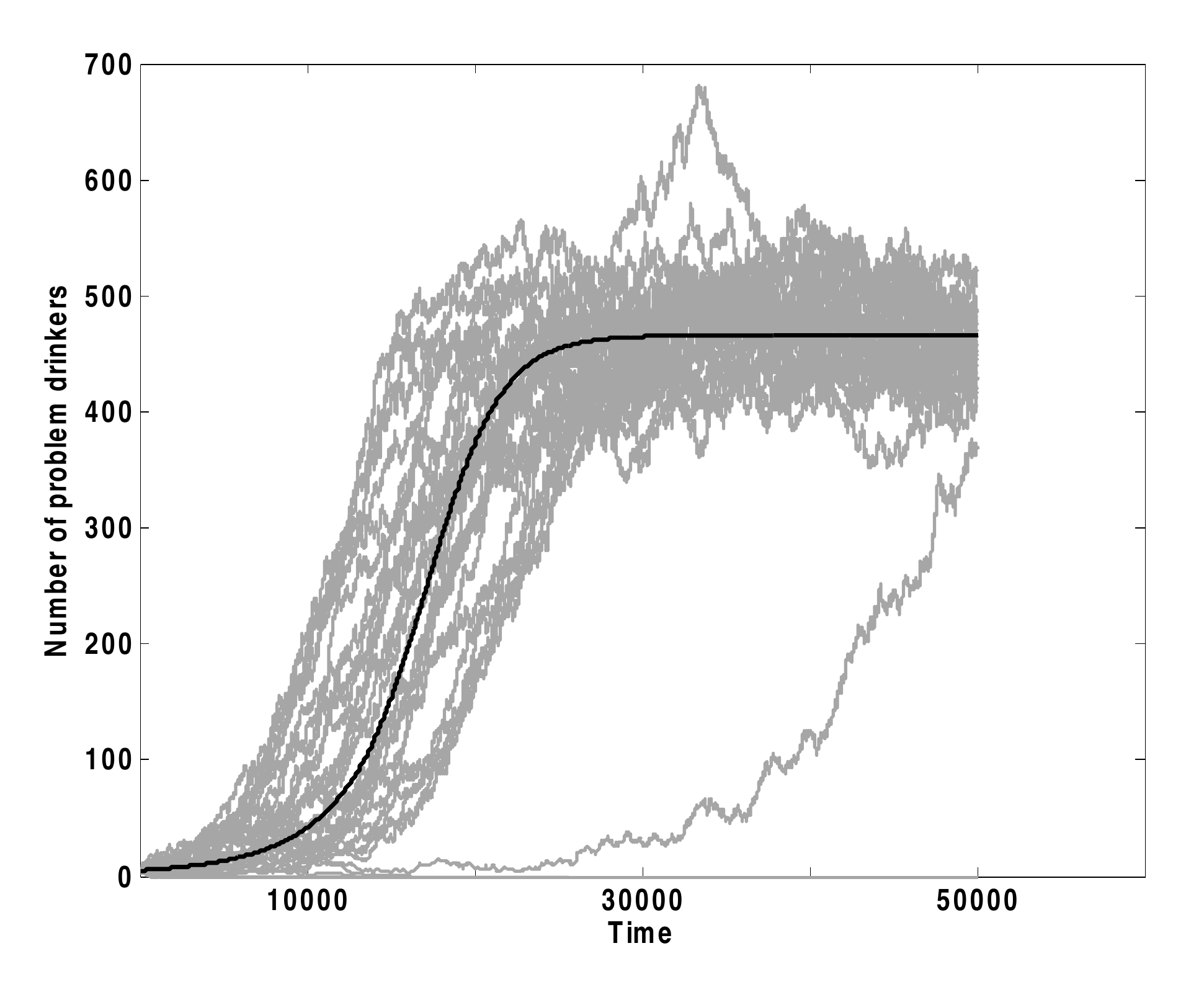}
        \caption{Results from numerical simulations. 50 stochastic
        realizations (grey curves) and numerical solutions of the deterministic (black curve)
	problem drinker class $D(t)$ versus time $t$.
          For these simulations the following values of parameters were used: $N=1000$, $\beta=1.20$, $\rho=7.00$, $\phi=0.50$
          and $\mu=0.50$ with ${\cal R}_\phi=1.20$ and the initial number of problem drinkers $D(0)=5$.}
        \label{stoch_sims}
    \end{center}
\end{figure}

The average behavior of the stochastic model is described in Table \ref{markov}. The simulations of this deterministic version and stochastic analog are computed using identical epidemiological and social parameter values. It is not surprising to see overall agreement between the dynamics of the deterministic model (black curve) and the mean (over 50 realizations) dynamics of the
stochastic model (grey curves) when ${\cal R}_{\phi}>1$ (Figure \ref{stoch_sims}). The mean results are computed under the condition of non-extinction of the $D$-class before the preselected time horizon. Setting  ${\cal R}_\phi<{\cal R}_c<1$ leads invariably to the eventual extinction of the $D$-class in the deterministic formulation but not always (as expected) in the stochastic formulation (Allen and van den Driessche 2006; Allen 2003).

In well-established drinking communities (including college students) estimates clearly show that ${\cal R}_\phi>1$.  Thus, one may ask whether the existence of backward bifurcations (bi-stability) is just of theoretical value?  If the goal is to prevent the formation of a drinking community then the above question ``makes" sense. However, most often the goal is to reduce or eliminate the $D$-class and the existence of a backward bifurcation makes this much harder.

Relapse rates among problem drinkers are high (Miller et al. 2001; Daido 2004). Hence, the existence of a relapse driven backward bifurcation suggests that efforts to ``eliminate"  problem drinkers or reduce problem drinking may be
futile as long as ``$R$-individuals" remain in the same social environment. Substantial reductions in the relapse parameter---with the ultimate goal of having ${\cal R}_{\phi}<1$---may be extremely difficult to achieve. Furthermore, treatment and prevention measures even if effective are likely to be insufficient if the goal is to eliminate the $D$-class (see bifurcation diagram in Figure \ref{alcofig}(a)).

\begin{figure}
    \begin{center}
	 \includegraphics[width=3in,height=2in]{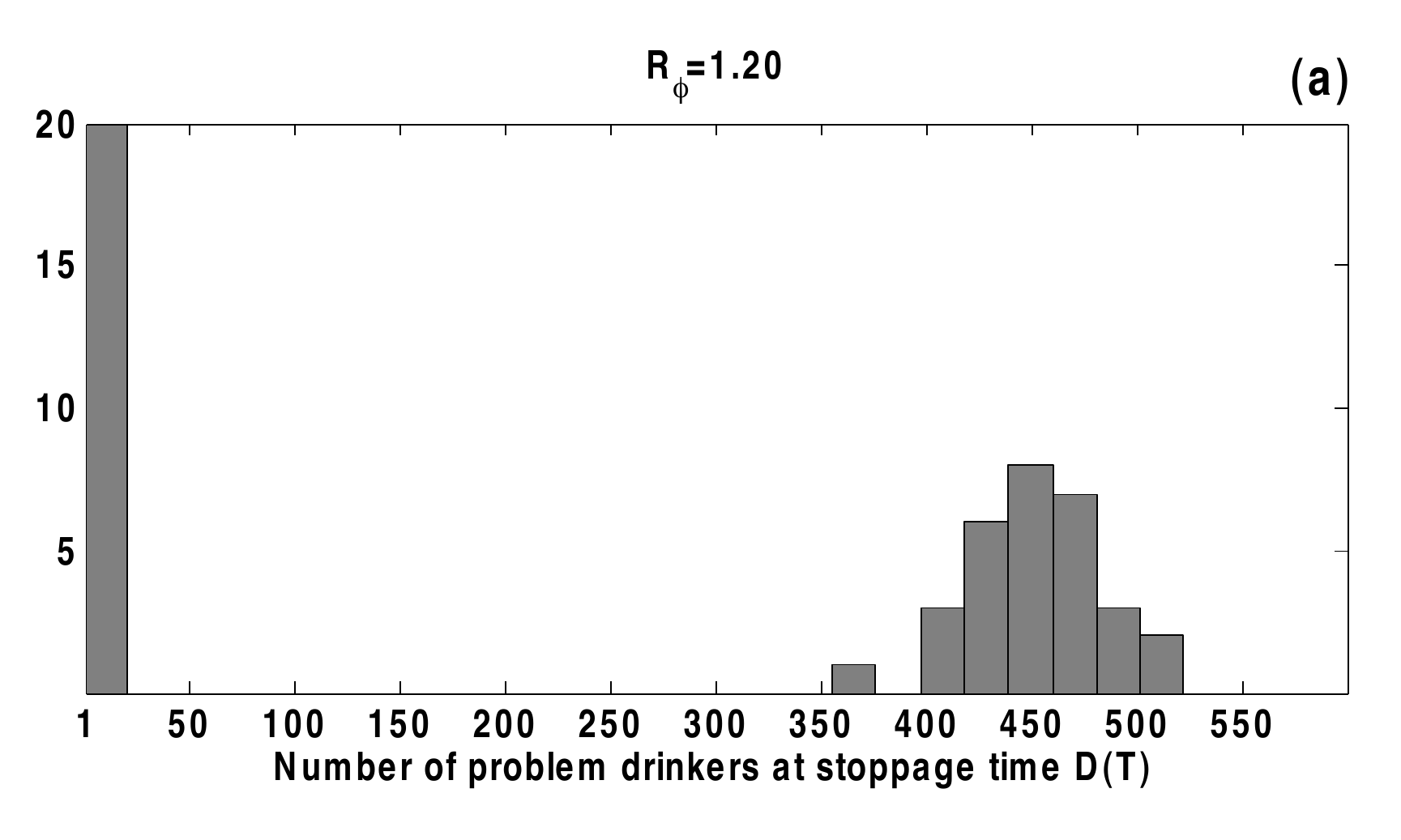}
	  \includegraphics[width=3in,height=2in]{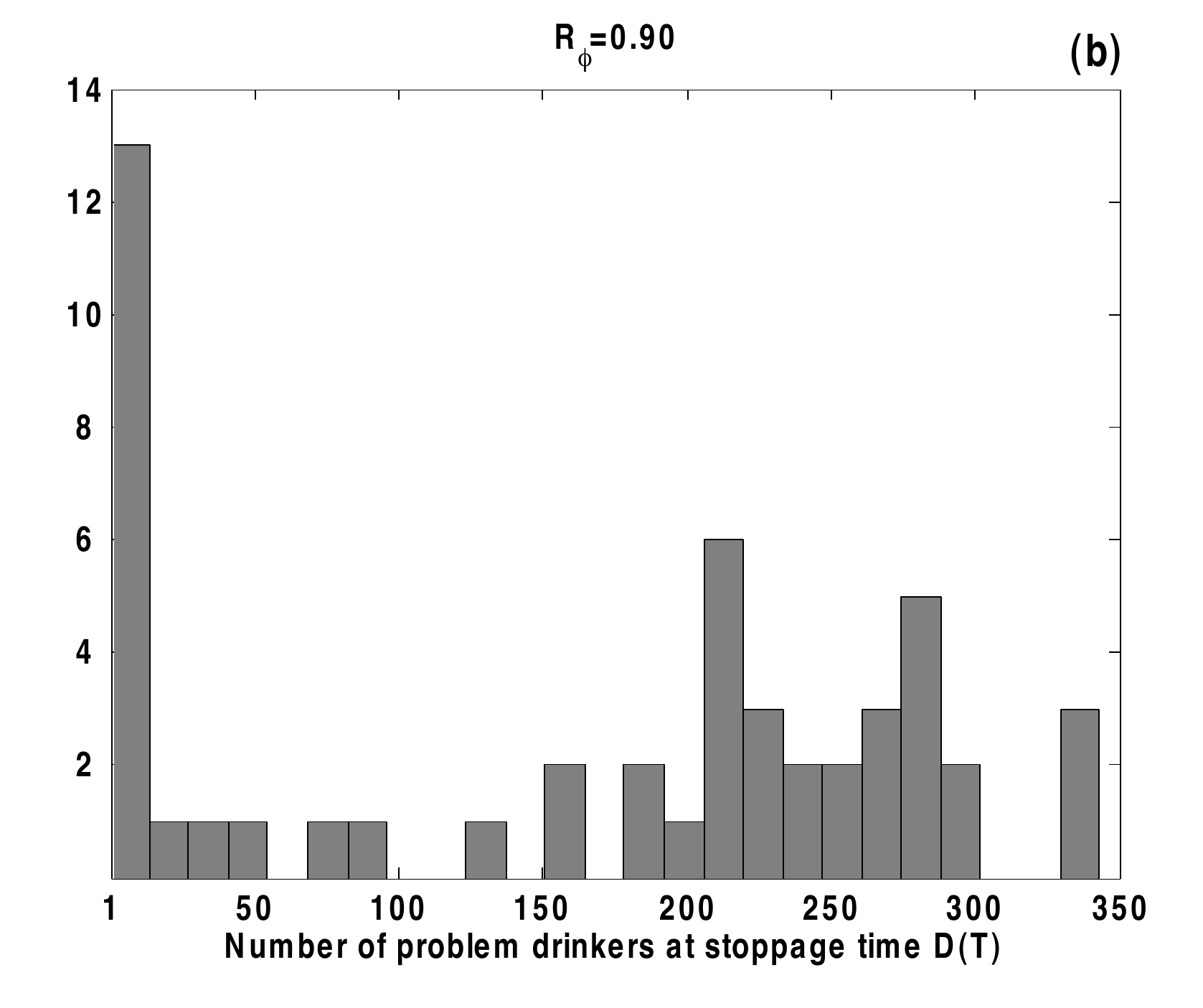}
        \caption{Histograms of $D(T)$, number of problem drinkers at stoppage time $T=50000$,
        resulting from 50 stochastic realizations with ${\cal R}_\phi>1$ (Panel (a))
        and $\mathcal{R}_{\phi}<1$ (Panel (b)).}
        \label{histd5}
    \end{center}
\end{figure}

Histograms (based on 50 stochastic realizations) of the number of problem drinkers at a stoppage time $T$,
denoted by $D(T)$, are examined when $\mathcal{R}_{\phi}>1$ (Figure \ref{histd5}(a)) and when $\mathcal{R}_{\phi}<1$
(Figure \ref{histd5}(b)).  Figure \ref{histd5}(a) shows that when ${\cal R}_\phi>1$ the value of $D(T)$ lies in $[350, 550)$ while Figure \ref{histd5}(b) shows that the problem drinker class may persist. Nearly forty percent
of the simulations involve result in a small segment of the population in the $D$-class (less than 10\%)  when ${\cal R}_\phi<1$ . These results are consistent with those of (S\'{a}nchez et al. 2007), that is, when the relapse rate is larger than the treatment rate ($\rho>\phi$). In other words, it is possible for a population of problem drinkers to become established even if ${\cal R}_\phi<1$ in a stochastic setting.

\section{Drinking Dynamics in Small-world Communities with High Relapse Rates}\label{swdrink}

 A network (graph) is a set of nodes with connections (edges)
between them.  Graphs provide visual representations of the contact structure of individuals in a population (Newman 2003). The fact that all social processes (including drinking) depend on contacts between
distinct individuals has, in part, motivated the study of epidemics on
networks (May and Lloyd 2001; Meyers et al. 2005; Pastor-Satorras and Vespignani 2001; Grabowski and Kosinski 2005).

Watts and Strogatz (Watts and Strogatz 1998) introduced a one-parameter, \emph{$p$}, family of networks. As the disorder parameter $p$ is varied in [0,1], the graph moves from a regular lattice
to a random graph.  The model can be formulated algorithmically as follows: the initial network is initialized via a one-dimensional periodic ring lattice of $N$ nodes, each connected to its
closest $\langle k \rangle$ neighbors (two nodes are neighbors if there is an edge connecting them).  The network
is updated by re-wiring each edge with probability $p$ (the disorder parameter) to a randomly
selected node until it reaches ``fixed" statistical properties. When $p\rightarrow0$ the algorithm recovers the initial lattice but when
$p\rightarrow1$, most edges are rewired, the resulting network is a random graph (Bollobas 2001).  Watts and Strogratz showed that the use of just a few random long-range connections ($p$ small) drastically reduced the \emph{average} distance between any pair of nodes (Watts and Strogatz 1998) ---the kind of property that enhances ``transmission'', the ``small-world effect''. The effect was postulated based on the result of a series of letter-forwarding experiments carried out by S. Milgram (Milgram 1967). The statistical properties of small-world and ``similar" networks have been studied (Watts and Strogatz 1998; Newman et al. 2006; and references therein).

Here we model community structure as a small-world network. The terms network and community are used interchangeably, with nodes representing individuals and edges denoting the
social connections or interactions, the kind of ``social mixing" that may lead to node ``transition'' (from the moderate drinker into the problem drinker state).  Nodes can be in one of three distinct states: moderate drinker, problem drinker, and recovered drinker.  The stochastic transitions between nodes' states are modeled as functions of time and the number of ``neighbors'' in particular states (transition rates).  If one starts with a community with $N$ nodes where Node $i$ ($1\leq i\leq N$) has $\delta(i,t)$ neighbors who, at time $t$, are in the state ``problem drinker'', then the probabilities that Node $i$ changes its state given that it alters its state, at each time step are: from moderate to problem drinker, $1-\exp(-\beta\delta(i,t))$; from problem to recovered, $1-\exp(-\phi)$; and from recovered to problem drinker, $1-\exp(-\rho_{\tau}(t)\delta(i,t))$.  This formulation (see Table \ref{sdr_prob}) defines a stochastic process on the random variables $S_p(t)$, $D_p(t)$, and $R_p(t)$. These random variables can also be thought of as parameterized by the disorder parameter $p \in [0,1]$.

%
%

\begin{figure}
        \centerline{\includegraphics[height=2.5in,width=4in]{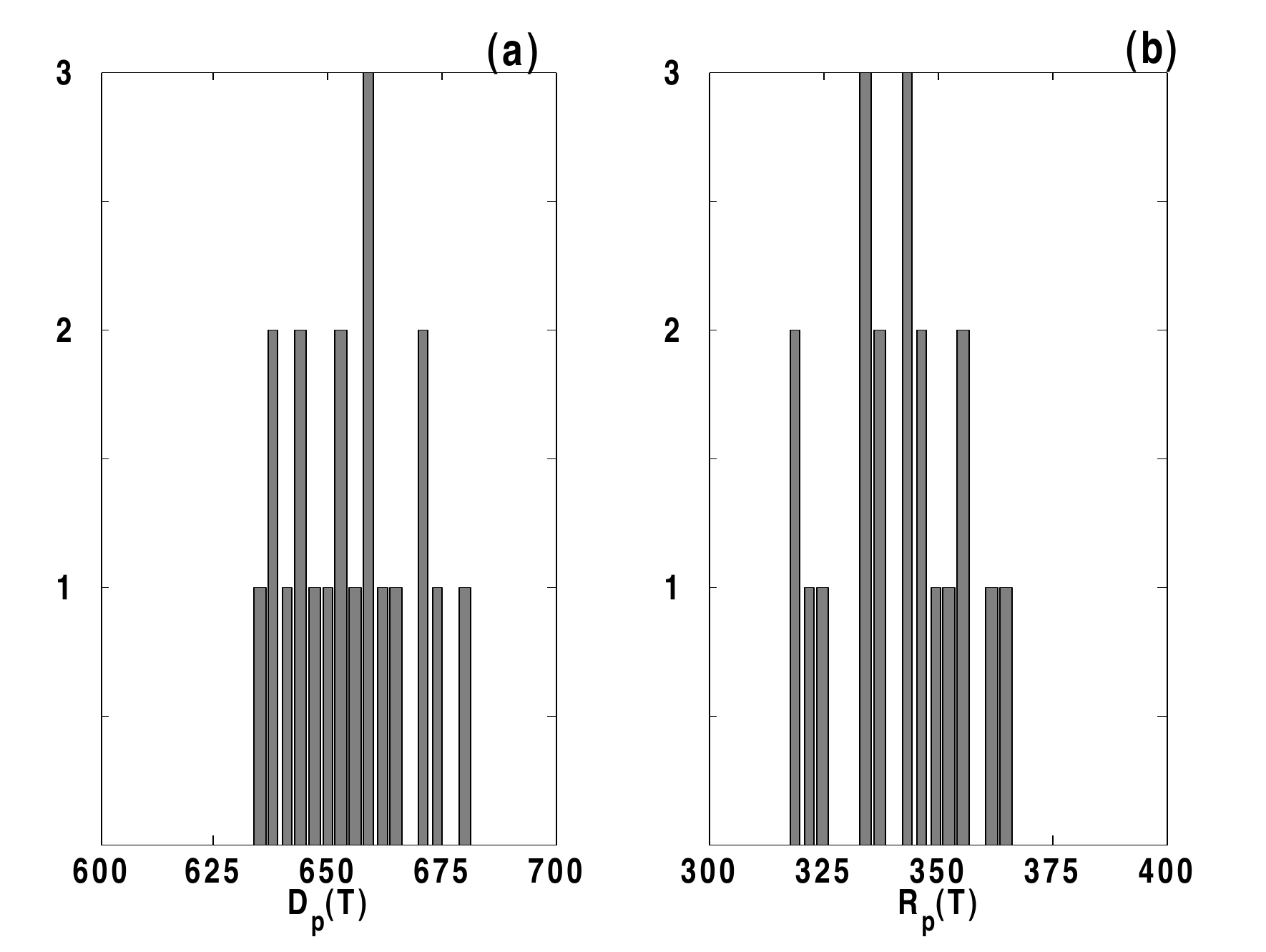}}
            \caption{Histograms of the total number of problem drinkers and
            recovered individuals, $D_p(T)$ and $R_p(T)$, respectively, at a
            stoppage time $T$.  Samples obtained from 20 stochastic realizations
        	   in simulated communities with $p= 3.02\times 10^{-4}$ in community size 1000 (nodes).}
            \label{rd_hists}
    \end{figure}
\begin{figure}
        \centerline{
            \includegraphics[height=3in,width=4in]{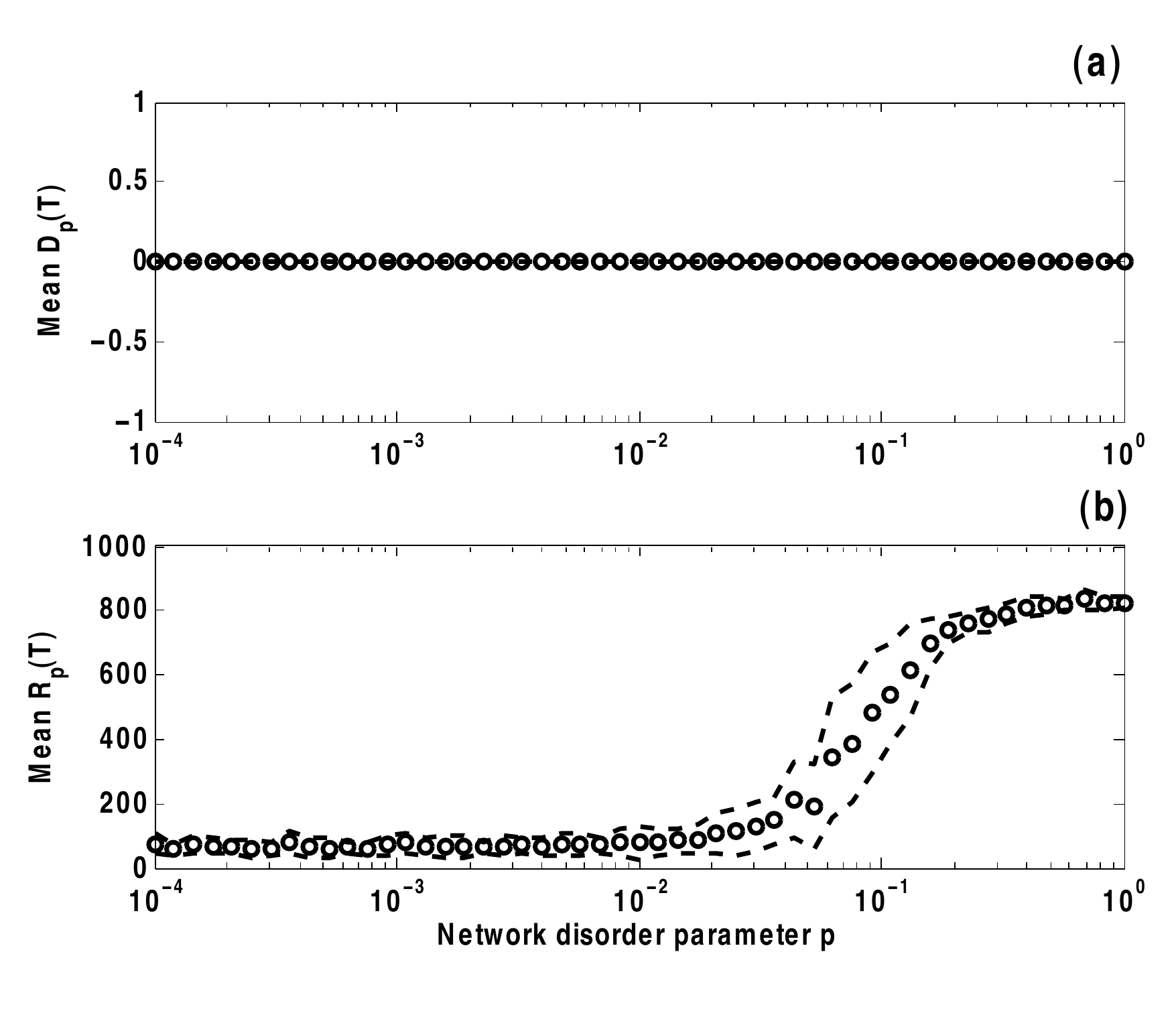}
            }
            \caption{Average and variance of $D_p(T)$ and $R_p(T)$ as functions of the simulated
        community architecture parameterized by $p$ (logarithmic scale).  The mean (circles)
        and mean plus and minus one standard deviation (dash curves) are computed from 20 stochastic
        realizations for each fixed value of $p$. Panels (a) and (b) display results of simulated
        contagion in small-world communities in the absence of relapse, $\rho\equiv0$.}
            \label{d_r_rhozero}
    \end{figure}
\begin{figure}
        \centerline{
            \includegraphics[height=3in,width=4in]{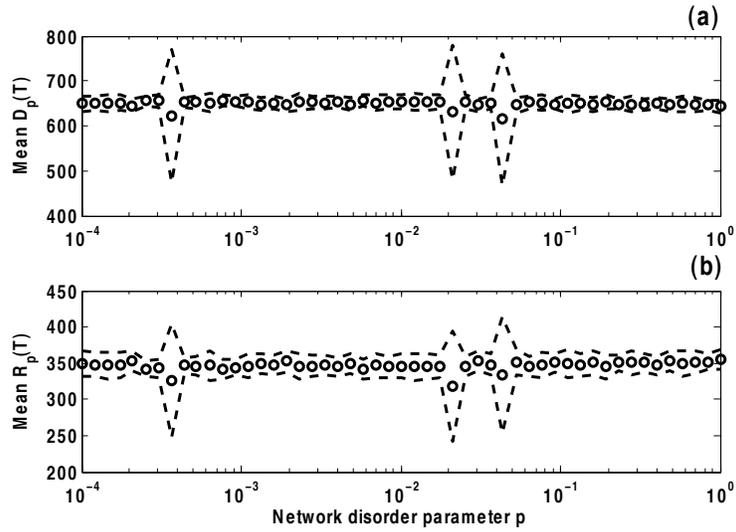}
            }
            \caption{Dependence of the average and variance of $D_p(T)$ and $R_p(T)$ as a function of community structure $p$ (logarithmic scale).  Average (circles) and
        one standard deviation added to and subtracted from the average (dash curves)
        are calculated from 20 stochastic realizations for each
        fixed value of $p$.  The results shown in Panels (a) and (b) assess a \lq\lq worst case scenario"
        of having on average every recovered node going into relapse with probability nearly one, in symbols
        $1-e^{-\rho_{\tau}(t)\langle k\rangle} \approx 1$.}
            \label{d_r_rho_worst}
    \end{figure}

Drinking as a ``contagious" process is simulated as follows: the stochastic generation of a small-world network (Watts and Strogatz 1998) is followed by multiple stochastic realizations of the drinking process defined in Table \ref{sdr_prob} on the selected small-world network. The parameter baseline values are summarized
in Table \ref{netwk_mdl_para}.  Histograms of $D_p(T)$ and $R_p(T)$, where $T$ denotes the stoppage time in
the simulations (see Table \ref{netwk_mdl_para}), are computed for each value of $p$ (see Figure \ref{rd_hists}). Figures \ref{d_r_rhozero} and \ref{d_r_rho_worst} highlight the mean and variance (over 20 realizations) of $D_p(T)$ and $R_p(T)$ as a function of $p$ (Chowell and Castillo-Chavez 2003; Chowell et al. 2006b).

A drinking wave is detected even as the size of the problem drinking class goes to zero for the case $\rho=0$ (no relapse) with ${\cal R}_\phi>1$. This feature agrees with deterministic (Brauer and Castillo-Chavez 2001) and stochastic ``theories" (Allen 2003) on single-outbreak SIR models. Figure \ref{d_r_rhozero}(a) shows that variations on the network structure (modeled by $p$) have no effect on the mean size of the problem drinker class
$D_p(T)$ . However, the mean size of the recovered class $R_p(T)$ exhibits a phase
transition as $p\rightarrow 10^{-1}$ (Figure \ref{d_r_rhozero}(b)).  Hence, in the absence of vital dynamics (births and deaths) and relapse, we conclude that community structure does affect the average size of the problem drinking class during the drinking wave. Small values of ``$p$'' lead to a phase transition (Newman 2003), a ``small world" effect.

Figure~\ref{d_r_rho_worst} illustrates a worst case scenario in which the average relapse probability is near one for the majority of the time. To see the impact of high, nearly stationary relapse rates, we let $\langle k\rangle$ denote the average number of connections per node
in a one-dimensional lattice when $p=0$ and carry out simulations on this network with the average relapse probability $(1-e^{-\rho_{\tau}(t)\langle k\rangle}) \approx 1$.  The relapse rate $\rho_{\tau}(t)$  (defined in Table \ref{netwk_mdl_para}) is modeled as a stepwise constant function that drops its value
at precisely $t=\tau$.  The worst case scenario here corresponds to the case where $\tau=\infty$. In general, when relapse rates are high for too long,  small-world structures (any value of $p$) have no effect on the mean sizes of the problem and recovered drinking classes. In fact, the size of the problem drinking community is above 60\% regardless of the value of $p$ (other parameters kept fixed).  Furthermore, we see that on average $D_p(T)+R_p(T)=N$ when relapse rates are high. That is, every member of this closed population becomes a problem drinker at least once regardless of the value of $p$.

\begin{figure}
        \centerline{
            \includegraphics[height=4in,width=6in]{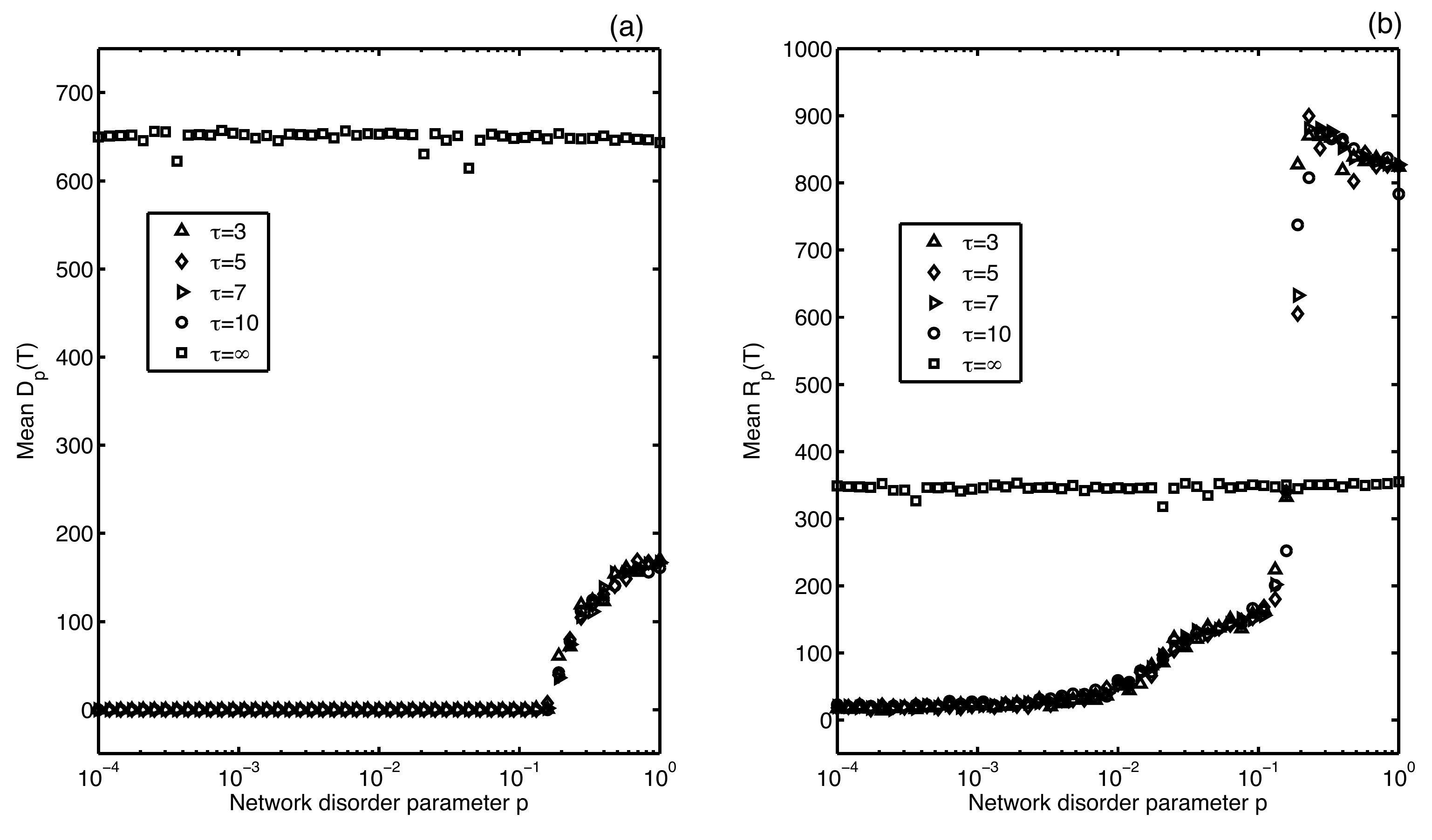}
            }
            \caption{Average $D_p(T)$ and $R_p(T)$ as functions of the community
        structure, $p$.  Panels (a) and (b) display the results
        obtained from using a time-dependent relapse rate $\rho_{\tau}(t)$.  The relapse
        rate jumps from 0.90 to 0.12 at time $t=\tau$, that is, every node diminishes its probability of transition from the recovered
        into the problem drinker  state by half (probabilities go from $1-e^{-0.90\langle k\rangle}\approx1$ to $1-e^{-0.12\langle k\rangle}\approx0.5$) Panels (a) and (b)  show the changes in averages as a function of the timing in the jump ($\tau$). The relapse reduction at times, $\tau=3$ (upward triangles), $\tau=5$ (diamonds),
        $\tau=7$ (right triangles), $\tau=10$ (circles) are highlighted.  The averages displayed in Figure
        \ref{d_r_rho_worst} are for the case $\tau=\infty$ (squares).}
            \label{d_r_rho_tau}
    \end{figure}

Reducing the relapse rate from 0.90 to 0.12 at precisely the time $\tau$ reduces the average relapse probability from $1-e^{-0.90\langle k\rangle}\approx1.00$
to $1-e^{-0.12\langle k\rangle}\approx0.50$ at time $\tau$.  Figure \ref{d_r_rho_tau} shows the impact of increasing the values of $\tau=3,\ 5,\ 7,\ 10$. We do not observe a lot of differences in the average values of $D_p(T)$ and $R_p(T)$ as a function of $\tau$. However, these averages ``improve" in the ``right'' direction as $\tau$ reduces its value from $\tau=\infty$ towards $\tau=0$.

\section{Discussion}\label{dissc}

Relapse has a significant impact on the dynamics of  addictive behavior (Gonzalez et al. 2003; S\'{a}nchez et al. 2007; Song et al. 2006; and references therein). The use of a simple system of differential equations (S\'{a}nchez et al. 2007) shows that for socially-intense processes like drinking,  the reproductive number, $\mathcal{R}_{\phi}$ is not always the key. Frequency dependent relapse rates play a huge role. Frequency dependent relapse rates do increase the possibility of severe outbreaks within ``well-behaved" communities, but more importantly they also increase the likelihood of failure of programs aimed at eliminating drinking. S\'{a}nchez et al. (S\'{a}nchez et al. 2007) clearly delineated the possibilities from their mathematical analysis of a simple model where all the mixing takes place in the same drinking environment. Mubayi et al. (Mubayi et al. 2008) recently explore the impact of individuals'  movement between heterogeneous drinking environments. They showed that frequent movement between \emph{distinct} environments can have a significant (negative) effect on the distribution of drinking types. Here, we only focused on exploring the predictions of (S\'{a}nchez et al. 2007) in two stochastic settings. The stochastic analog (continuous time Markov chain) of S\'{a}nchez et al.'s deterministic model was used to highlight the role of variability. The results were consistent with those of Sanchez et al. with the usual caveats (Allen 2003). A small-world network was used to highlight the very strong role played by relapse.

In fact, our study of drinking in a small-world network parameterized by the disorder parameter $p$ leads to the following results: When there is no relapse ($\rho=0$), we recovered the well understood phase transition effect previously identified from SIR simulations on small-world networks (Newman 2003),
as $p$ crosses a critical value; the introduction of high relapse rates ``eliminates'' the role of ``$p$''. In other words, the form of social connections (who interacts with whom) in populations experiencing strong patterns of relapse has no impact on the prevalence of addictive behaviors. Hence, if relapse rates are high then emphasis on programs that generate substantial and sustained reductions in ``mixing" will not be effective. Reducing residence times in risky environments which promote relapse, reducing recruitment into drinking communities and reducing movement between drinking venues are more likely to be effective (Mubayi et al. 2008).

\section{Appendix}
	Transitions between drinking classes involve discrete events which
	change the number of individuals in every class, one at a time.  For example,
	when a drinking ``contagion'' event occurs, the number of moderate drinkers is
	decreased by one, while the number of problem drinkers increases by one.  The
	probability that an event takes place during an infinitesimal time interval $[t, t+dt]$
	is calculated from the average rates in the deterministic model.  In this example, the ``conversion'' event occurs at the rate
	of $\beta S(t) D(t)/N$ and the probability that it happens in $[t,t+dt]$ is approximately $\left(\beta S(t) D(t)/N\right) dt$.  All the
	events, their rates of occurrence, and the probabilities at which they take place are listed in Table \ref{markov}.
	
	It is assumed that the events are described by independent Poisson
	processes (Allen, 2003).  The term
		\[E=\mu N+ \mu S+ \mu D+ \mu R +\beta S D/N+ \phi D+ \rho R D /N,\]
	denotes the rate at which an event occurs at time $t$.  The time between events is exponentially
	distributed with mean $1/E$. The time at which the next event happens is found, for each realization, by sampling
	from an exponential distribution with mean $1/E$.
	
	To decide which event takes place (once it is known that an event occurs), we divide up the interval
	$(0,E)$ into subintervals that correspond to the relative occurrence probabilities of the various events.  For example, given
	that an event has occurred, the probability that it is a recruitment is $\mu N/E$,
	the probability of the removal of a moderate drinker is $\mu S/E$, the probability of the removal of
	a problem drinker is $\mu D/E$, etc.  A number $U$ is selected randomly from the uniform
	distribution on $(0,1)$ and an event is selected if this value falls within the appropriate subinterval. For instance,
	the event is a recruitment if $U$ satisfies $0<U<\mu N/E$, a moderate drinker removal if $U$ lies between $\mu N/E$ and $(\mu N+\mu S)/E$,
	a problem drinker removal if $U$ lies between $(\mu N+\mu S)/E$ and $(\mu N+\mu S + \mu D)/E$, and so on.

\section*{Acknowledgments}
A.C.-A. was supported in part by the Statistical and Applied Mathematical Sciences Institute which is
funded by the National Science Foundation under Agreement No. DMS-0112069. Any opinions, findings, and
conclusions or recommendations expressed in this material are those of the authors and do not necessarily
reflect the views of the National Science Foundations.  A.C.-A.  was also supported in part by Grant Number
R01AI071915-07 from the National Institute of Allergy and Infectious Diseases.  The content is solely the responsibility
of the authors and does not necessarily represent the official views of the NIAID or the NIH. X.W. and C.C.-C. were supported by NIAAA grant on \lq \lq Ecosystem Models of Alcohol-Related Behavior", Contract No. HHSN2S1200410012C, ADM Contract No. No1AA410012 through Prevention Research Center, PIRE, Berkeley, the National Science Foundation (DMS-0502349), the National Security Agency (DOD-H982300710096), the Sloan Foundation, and Arizona State University. P.J.G. was supported by Grant Number R01 AA06282 from the National Institute on Alcohol Abuse and Alcoholism.

\clearpage

\section{Tables}
\begin{table}[h]
    \caption{State variables and parameters of the contagion model in (S\'{a}nchez et al. 2007).}
    \begin{center}
        \begin{tabular}{|c|l|} \hline\hline
            State variable& Description\\ \hline\hline
            $S(t)$ & Number of occasional and moderate drinkers at time $t$ \\ \hline
            $D(t)$ & Number of problem drinkers at time $t$\\ \hline
            $R(t)$ & Number of recovered individuals at time $t$\\ \hline \hline
            Parameter& Description\\ \hline\hline
            $\beta$& Effective transmission rate (average number of effective interactions   \\
            & per occasional and problem drinker per unit of time)\\ \hline
            $\rho$ & Community-driven relapse rate (average number of effective interactions  \\
            & per problem drinker and recovered individual per unit of time) \\ \hline
            $\phi$& Per-person treatment rate\\ \hline
            $\mu$ & Per-person departure rate from the drinking environment\\ \hline
            $N$ & Community size (permanent population size) \\ \hline
        \end{tabular}
    \end{center}
    \label{ode_mdl_vars}
\end{table}

\begin{table}[h]
    \caption{Collects the transition rates and infinitestimal probabilities of occurrence of the events linked to a single drinking model outbreak. The dependence on $t$ is omitted, writing $S$, $D$, and $R$,
    instead of $S(t)$, $D(t)$, and $R(t)$, respectively.}
    \begin{center}
        \begin{tabular}{llcc}\hline\hline
            Event& Transition & Rate at which & Probability of transition \\
            & & event occurs& in time interval $[t, t+dt]$ \\ \hline\hline
            Recruitment &$S\rightarrow S+1$ & $\mu N$ & $\mu N dt$ \\
            Moderate drinker removal &$S\rightarrow S-1$ & $\mu S$ & $\mu S dt$ \\
            Problem drinker removal &$D\rightarrow D-1$ & $\mu D$ & $\mu D dt$ \\
            Sober removal &$R\rightarrow R-1$ & $\mu R$ & $\mu R dt$ \\
            Drinking contagion &$S\rightarrow S-1$, $D\rightarrow D+1$ & $\beta S \frac{D}{N}$ & $\beta S \frac{D}{N}dt$\\
            Recovery & $D\rightarrow D-1$, $R\rightarrow R+1$& $\phi D$ & $\phi D dt$\\

            Relapse & $D\rightarrow D+1$, $R\rightarrow R-1$& $\rho R \frac{D}{N}$ & $\rho R \frac{D}{N} dt$\\
        \end{tabular}
    \end{center}
    \label{markov}
\end{table}

\begin{table}
    \caption{State variables, parameters, events, and transition probabilities of the drinking dynamics model in small-world communities.}
    \begin{center}
        \begin{tabular}{|c|l|} \hline\hline
            State variable&  Description \\ \hline\hline
            $\delta(i,t)$& Number of problem drinker neighbors of node $i$ at time $t$\\ \hline
            $S_p(t)$& Total number of moderate drinkers at time $t$ in a \\
            &small-world community parameterized by $p$ \\ \hline
            $D_p(t)$&Total number of problem drinkers at time $t$ in a \\
            &small-world community parameterized by $p$ \\ \hline
            $R_p(t)$& Total number of recovered individuals at time $t$ in a \\
            &small-world community parameterized by $p$ \\ \hline\hline
             Parameter& Description \\ \hline\hline
            $\beta$& Transmission rate\\ \hline
            $\phi$ & Per-person treatment rate\\ \hline
            $\rho_{\tau}(t)$ & Time-dependent relapse rate\\ \hline\hline
        \end{tabular}
\begin{tabular}{lc}\hline\hline
            Event &  Probability of transition\\ \hline\hline
            Node $i$ changes from  {\em moderate} into {\em problem} drinker & $1-e^{-\beta\delta(i,t)} \) \\ \hline
            Node $i$ switches from {\em problem drinker} into {\em recovered}& $1-e^{-\phi}$\\ \hline
            Node $i$ changes from {\em recovered} into {\em problem drinker} & $1-e^{-\rho_{\tau}(t)\delta(i,t)}$\\ \hline
\end{tabular}
    \end{center}
    \label{sdr_prob}
\end{table}

\begin{table}
    \caption{Parameter values utilized in simulations of drinking dynamics in small-world communities.}
    \begin{center}
        \begin{tabular}{|c|l|c|} \hline\hline
            Parameter &Description & Baseline value \\ \hline\hline
            $\langle k\rangle$& Average connectivity per node& 6 \\ \hline
            $N$& Community size&1000 \\ \hline
            $\beta$& Transmission rate& 0.12\\ \hline
            $\phi$&Per-person treatment rate& 0.7\\ \hline
            $\rho_{\tau}(t)$& Time-dependent relapse rate&
            $\rho_{\tau}(t)= 0.90$ whenever $t<\tau$\\
            &&$\rho_{\tau}(t)= 0.12 $ if $t\geq\tau$ \\ \hline
            $T$&Stoppage time& 4000 \\ \hline
            $D_p(0)$&Initial number of problem drinkers chosen &\\
            &uniformly at random in every community&5 \\ \hline
            &Number of stochastic realizations& 20\\ \hline
        \end{tabular}
    \end{center}
    \label{netwk_mdl_para}
\end{table}

\clearpage

\section{Figures}	

\end{document}